\documentclass[prd,twocolumn,nofootinbib]{revtex4}
\usepackage{bm,amsmath,amssymb,graphicx}

\begin{document}

\title{NON-PARTICLE DARK MATTER FROM HUBBLE PARAMETER}
\author{Nikodem J. Pop{\l}awski}
\affiliation{Department of Mathematics and Physics, University of New Haven, West Haven, CT, USA}
\email{NPoplawski@newhaven.edu}

\noindent
{\em The European Physical Journal C}\\
Vol. {\bf 79}, No. 9, p. 734 (2019)
\vspace{0.4in}

\begin{abstract}
The measurements of the Hubble parameter using the cosmic microwave background radiation appear to be inconsistent with the measurements of this parameter using Cepheid variable stars.
This inconsistency may be a result of using the $\Lambda$CDM cosmology, which assumes pressureless dark matter, in extrapolating the data from the recombination time to the present time.
We show that both measurements are consistent if dark matter satisfies an equation of state in which the pressure $p$ and the energy density $\epsilon$ are related by $p=w\epsilon$ with a negative value of $w$.
The data give $w\approx -0.01$.
The negative value of $w$ indicates that dark matter would not be formed by particles, which is consistent with the lack of experimental evidence for them.
\end{abstract}
\maketitle

The observations of the temperature angular power spectrum of the cosmic microwave background (CMB) radiation provide the information about the composition of the universe \cite{peaks}.
The measurements of the position of the first acoustic peak show that the universe is nearly flat, with the density parameter $\Omega\approx 1$.
The measurements of the amplitudes of the peaks determine the values of $\Omega_\textrm{b}h^2$ and $\Omega_\textrm{c}h^2$, where $\Omega_\textrm{b}$ is the density parameter for baryonic matter, $\Omega_\textrm{c}$ is the density parameter for dark matter, and the present-day Hubble parameter $H_0=100h$ km/s/Mpc.
Combining these values with the measured $h=0.674\pm 0.005$ gives (omitting errors) $\Omega_\textrm{b}=0.049$ and $\Omega_\textrm{c}=0.265$, as obtained by the Planck satellite \cite{Planck}.
Consequently, the density parameter for total matter is $\Omega_\textrm{m}=0.314$.

Before the Planck measurements, most measured values of $h$ also clustered around $0.68$, including the data from high-redshift type Ia supernovae (SN Ia) \cite{val1}, Wilkinson Microwave Anisotropy Probe CMB data \cite{val2}, baryon acoustic oscillations (BAO) \cite{val3}, SN Ia and BAO data \cite{val4}, measurements of the Hubble parameter at intermediate redshifts \cite{val5}, and large-scale-structure data \cite{val6}.
Assuming a flat universe, $\Omega_\Lambda=1-\Omega_\textrm{m}=0.686$ for dark energy (cosmological constant).
The resulting age of the universe is $t_\textrm{univ}=\int da/(aH)=\int_0^\infty dz/[H_0(1+z)\sqrt{\Omega_\textrm{m}(1+z)^3+\Omega_\Lambda}]$, where $H=\dot{a}/a$ is the Hubble parameter as a function of the scale factor $a$ and its time derivative $\dot{a}$, and $z$ is the redshift \cite{Ric}.
The numerical values give $t_\textrm{univ}=13.8 \times 10^9$ years.

The observations of Cepheid variable stars in the Large Magellanic Cloud provide a different value of the present-day Hubble parameter, $\tilde{h}=0.742\pm 0.018$ \cite{Ceph}.
A tilde denotes a value measured locally.
Moreover, the observations of quasars gravitationally lensed by galaxies give even a larger value \cite{qua}.
This apparent discrepancy between the CMB and Cepheid measurements cannot be attributable to an error and has caused the so-called "Hubble tension", suggesting physics beyond $\Lambda$CDM \cite{tens}.
Other local measurements of the expansion rate have found a lower value with a larger uncertainty, including the data from SN Ia \cite{other1}, low-redshift SN Ia \cite{other2}, and the ionized gas in HII galaxies \cite{other3}.

Many scenarios to resolve the Hubble tension have been proposed, including modified dark energy \cite{daen} and decaying dark matter particles \cite{dec}.
In this note, we argue that Hubble tension may be a result of extrapolating the CMB data from the time of recombination ($z=1089$) to the present time ($z=0$) using the $\Lambda$CDM cosmology, which assumes pressureless dark matter.
We use the cosmological constant as dark energy because it can naturally arise from the simplest Lagrangian for the gravitational field in which the torsion tensor is the only variable \cite{affine}.
We show that dark matter satisfying an equation of state $p=w\epsilon$ \cite{eos}, where $p$ is the pressure and $\epsilon$ is the energy density, with $w\approx -0.01$ instead of $w=0$ removes the discrepancy.

The CMB data do not measure directly $\Omega_\textrm{b}$ but rather $\Omega_\textrm{b}h^2(1+z)^3$, where $z=1089$.
Therefore, a different value of $h$ yields a different value of $\Omega_\textrm{b}$.
The two interpretations of the same result require
\begin{equation}
\Omega_\textrm{b}h^2(1+z)^3=\tilde{\Omega}_\textrm{b}\tilde{h}^2(1+z)^3,
\end{equation}
from which we obtain the real value of the density parameter for baryonic matter, $\tilde{\Omega}_\textrm{b}=0.040$.
Similarly, the CMB data do not measure directly $\Omega_\textrm{c}$ but rather $\Omega_\textrm{c}h^2(1+z)^3$.
Therefore, a different value of $h$ yields a different value of $\Omega_\textrm{c}$.
In addition, dark matter may have a nonzero value of $w$ and thus scale with $z$ as $(1+z)^{3(1+w)}$ \cite{Ric}.
For dark matter, we therefore require
\begin{equation}
\Omega_\textrm{c}h^2(1+z)^3=\tilde{\Omega}_\textrm{c}\tilde{h}^2(1+z)^{3(1+w)}.
\end{equation}
This equation means that the CMB data must be extrapolated to the present time, in order to determine the value $\tilde{\Omega}_\textrm{c}$, using the equation of state with $w$.
Combining this equation with $\tilde{\Omega}_\textrm{b}+\tilde{\Omega}_\textrm{c}=1-\Omega_\Lambda$ (for a flat universe), which gives the real value of the density parameter for dark matter, $\tilde{\Omega}_\textrm{c}=0.274$, determines the value of $w$:
\begin{equation}
w=-0.0108.
\end{equation}
The value of $w$ is negative because $\tilde{\Omega}_\textrm{c}>\Omega_\textrm{c}$ and $\tilde{h}>h$.
This negative value indicates that dark matter is not formed by particles, for which $w$ must lie between 0 (nonrelativistic limit) and 1/3 (ultrarelativistic limit) \cite{LL2}.

This value agrees with the results of \cite{eos}, which analyzed the dark matter equation of state between $z=10^5$ and $z=0$ using the CMB and BAO data.
Those results showed that $w$ is consistent with 0 but allows a small negative value.
This value also agrees with the results of \cite{clus}, which analyzed the dark matter equation of state by measuring galaxy cluster mass profiles.
That analysis reported $w=0.00\pm 0.15 \pm 0.08$.

The resulting age of the universe is
\begin{equation}
t_\textrm{univ}=\int_0^\infty \frac{dz/(1+z)}{H_0\sqrt{\tilde{\Omega}_\textrm{b}(1+z)^3+\tilde{\Omega}_\textrm{c}(1+z)^{3(1+w)}+\Omega_\Lambda}}.
\end{equation}
The numerical values give $t_\textrm{univ}=12.7 \times 10^9$ years.

If dark matter were a kind of fluid, it would have to satisfy the laws of fluid mechanics.
For a barotropic equation of state $p=w\epsilon$, where $w$ is constant, the adiabatic speed of sound in the fluid is equal to $c_s=\sqrt{\partial p/\partial \epsilon}c=\sqrt{w}c$, where $c$ is the speed of light \cite{LL6}.
For a negative $w$, the speed of sound is imaginary and the dark matter fluid would be unstable and rapidly become spatially inhomogeneous.
To avoid this instability, one must then specify an additional condition $c_s=c$, as for scalar fields \cite{DE}.
In this case, however, dark matter would not cluster on galactic scales.
Also, the observed flat rotation curves of spiral galaxies constrain the speed of sound of dark matter to be $c_s<10^{-4}c$ \cite{sound}.

Therefore, if dark matter has a generalized equation of state \cite{eos,gen} with a negative parameter $w$, such a parameter has no fluid-mechanical interpretation; it only represents how dark matter interacts gravitationally and influences the dynamics of the universe.
Accordingly, dark matter (without additional generalizations) cannot be a dark fluid composed of particles and its non-particle nature should be investigated.
This conclusion was also reached in \cite{univ} using the Einstein--Cartan theory of gravity \cite{EC,cosmo}.
This conclusion agrees with the experimental lack of evidence for various hypothetical particles that have been proposed as candidates for dark matter, such as axions \cite{axion} or weakly interacting massive particles \cite{wimp} that are predicted by supersymmetric scenarios \cite{susy}.

This work was funded by the University Research Scholar program at the University of New Haven.

\end{document}